\documentstyle[12pt]{article}
\begin{document}
\begin{titlepage}
\pagestyle{empty}
\begin{flushright}
\end{flushright}

\vskip 0.5cm

\begin{center}
{\large \bf Cosmology, Thermodynamics and Matter Creation}
{\footnote[1]{This paper was originally published in Frontier Physics, Essays in Honor of Jayme Tiomno, Eds. S. Mac Dowel, H. M. Nussenzweig and R. A. Salmeron, World Scientific, Singapure, p. 317-326, 1991.}}
\vskip 0.5cm
{\large \bf (In Honor of Prof. Jayme Tiomno)}
\vskip 0.5cm
\renewcommand{\thefootnote}{\alph{footnote}}
J. A. S. Lima$^{}$\footnote{e-mail:limajas@astro.iag.usp.br}
\vskip 0.2cm
\begin{center}
{  Universidade  
Federal do Rio Grande do Norte (UFRN) \\Departamento de F\'{\i}sica Te\'{o}rica e Experimental, 
    C. P. 1641\\
     59072 - 970, Natal, RN, Brazil}
\end{center}
\vskip 0.2cm
{M. O. Calv\~ao$^{}$\footnote{e-mail:orca@if.ufrj.br} and I. Waga$^{}$\footnote{e-mail:iwaga@if.ufrj.br}}
\end{center}
\begin{center}
\vskip 0.2cm
{Universidade Federal do Rio de Janeiro (UFRJ)\\ Instituto de F\'{\i}sica, Ilha do Fund\~ao, C. P. 68528\\ 21945, Rio de Janeiro, RJ, Brazil}
\end{center}

\vskip 1.0cm

\begin{abstract}
\vskip 1.0cm
\noindent Several approaches to the matter creation problem in the context of cosmological models are summarily reviewed. A covariant formulation of the general relativistic imperfect simple fluid endowed with a process of matter creation is presented. By considering the standard big bang model, it is shown how the recent results of Prigogine et alii \cite{1} can be recovered and, at the same time their limits of validity are explicited. 

\vskip 1.0cm

\end{abstract}
\end{titlepage}

\pagebreak

\baselineskip 0.6cm

\section{Introduction}

The study of matter creation processes in the context of the
cosmological models for the universe is fairly old. Seemingly,
there is a generalized belief that if the mass of the particles or
the number of particles itself are time-dependent functions, the
scale of time driving such processes could, in principle, be
established only in the domain of a given cosmological model. In
this sense, the matter creation phenomenon is, from the very
beginning, closely related to the cosmological problem.

In the steady-state model \cite{2}, for instance, the observed cosmic
expansion, together with the perfect cosmological principle, leads
necessarily to a continuous creation of matter. In this model the
energy conservation law seems to be saved by adding a new term
(C-field) to the standard Einstein-Hilbert action. Also, after
Dirac \cite{3}, in the majority of variable gravity theories, the time
variation of the gravitational constant has as a consequence a
time-dependence of the mass and/or the particle number.

Recently, following a quite different approach, Wesson \cite{4} proposed a
Kaluza-Klein-type variable-mass gravity theory. He argued that
just as the speed of light $c$ is used to define $x^o = ct$,
putting time in an equal footing with the space coordinates, one
can consider the gravitational constant $G$ to define a new
coordinate $x^4 = Gm/c^2$ and use it as the fifth coordinate of a
five-dimensional space-time-mass (manifold). In this case, the
variation rate of the mass will be related to the ``velocity"
$\frac{dx^4}{dx^o}= \frac{G}{c^3}\frac{dm}{dt}$ of the particle in
the fifth dimension. However, some simple arguments developed by
Wesson himself strongly suggest that the mass dimension is again
important only in the realm of cosmology.

At first sight, the several approaches mentioned above seem to be
quite different, but, to a certain extent, there is a unifying
route among them, namely, only the dynamic aspects of the matter
creation process have been investigated.

More recently, Prigogine and Collaborators \cite{1} taking advantage
of the thermodynamic theory of irreversible processes, dealt with
the matter creation problem in the framework of the standard hot
big bang cosmology. The main goal of their paper was to construct
a coherent phenomenological approach in which both matter and
space-time curvature were simultaneously generated. In this case,
as originally envisaged by Tryon \cite{5} the singularity can be
avoided, with the universe now observed by us arising from a kind
of instability of Minkowski space-time.

In the present work we reexamine the main results of Prigogine
{\it et al.}'s paper from a more general point of view.
Particularly, it will be shown that, if the specific entropy (per
particle) is not constant, several of their results are no longer
valid.

\section{Thermodynamics and Matter Creation for a Simple Fluid}

The basic macroscopic variables describing the thermodynamic
states of a relativistic simple fluid are the energy-momentum
tensor $T^{\alpha \beta}$, the particle flux vector $N^{\alpha}$,
and the entropy flux vector $s^{\alpha}$.

We will restrict our considerations to an energy-momentum tensor
of the form

\begin{equation}
T^{\alpha \beta} = (\rho + P)u^{\alpha}u^{ \beta}  - P g^{\alpha
\beta},
\end{equation}
satisfying the conservation law

\begin{equation}
T^{\alpha \beta}_{; \beta}=0.
\end{equation}

In (1), $\rho$ is the energy density, $P$ is the (isotropic)
dynamic pressure, and $u^{\alpha}$ is the fluid four-velocity.
Here and henceforth we assume spatial isotropy.

The dynamic pressure $P$ may be decomposed as
\begin{equation}
P = p + \prod
\end{equation}
where $p$ is the equilibrium (thermostatic) pressure and $\prod$
is a correction term present in dissipative situations.

The particle flux vector will be assumed to have the form

\begin{equation}
N^{\alpha} =n u^{\alpha},
\end{equation}
where $n$ is the particle density. It satisfies the balance law
\begin{equation}
N^{\alpha}_{; \alpha}= \Psi ,
\end{equation}
where $\Psi$ is a particle source ($\Psi > 0$) or sink ($\Psi <
0$) term. Usually in cosmology it is taken equal to zero.

The entropy flux vector is supposed to be
\begin{equation}
S^{\alpha} =n\sigma u^{\alpha},
\end{equation}
where $\sigma$ is the specific entropy (per particle). It
satisfies the second law of thermodynamics

\begin{equation}
S^{\alpha}_{;\alpha}\geq 0.
\end{equation}

For this system, the Gibbs relation reads

\begin{equation}
nT d\sigma = d\rho - \frac{\rho + p}{n} dn,
\end{equation}
where $T$ is the temperature.

From the equations above, it is easy to show that

\begin{equation}
S^{\alpha}_{;\alpha} = - \frac{\prod \Theta}{T} -
\frac{\mu\Psi}{T} ,
\end{equation}
where $\Theta = u^{\alpha}_{;\alpha}$ is the expansion of the
fluid, and $\mu$ is its chemical potential defined by Euler's
relation
\begin{equation}
\mu = \frac{\rho + p}{n} - T\sigma .
\end{equation}

Eq. (9), as far as we know, is presented here for the first time
in the literature.

When the particle number is conserved ($\Psi = 0$), $\Pi$ stands
for the bulk viscosity of the fluid. This kind of viscosity occurs
in almost every situation. For instance, it is a well-know result
from kinetic theory that a relativistic simple gas of weakly
interacting point particles does not expand adiabatically. We can
account for this phenomenon if we conceive the gas as a mixture of
two components with different specific heats, each one expanding
in an adiabatic manner during a time of the order of the mean free
time. Each component will cool down at a different rate, thus
producing a kind of ``microscopic temperature gradient" over
distances of the order of the mean free path. The heat flux
tending to reequalize the temperatures is the mechanism of
dissipation, which, in this context, is identified with the bulk
viscosity \cite{6}.

Besides the usual classical meaning described above, the ``viscous
pressure" $\Pi$ is also relevant when there occur processes in
which a variation of the total number of particles takes place
($\Psi \neq 0$). In this case, it is commonly denoted as a
``creation pressure". For instance, in an expanding universe in
which there is matter creation, we must expect that the energy in
a comoving volume decreases more slowly than in the ordinary
situation. This property is taken account of by means of a
reduction in the effective dynamic pressure. Formally, Eq. (2)
implies that

\begin{equation}
\frac{d(\rho \Delta V)}{dt} = - P\frac{\Delta V}{dt} ,
\end{equation}

Assuming $\frac{\Delta V}{dt} > 0$ (expansion), we must have $P <
p$, so that $\rho \Delta V$ falls more slowly than in equilibrium.

Both processes described above are scalar ones and may take place
simultaneously. However, from now on, we will restrict ourselves
to the process of matter creation only. We shall suppose that the
particles spring up into space-time in such a way that they turn
out to be in thermal equilibrium with the already existing ones.
The entropy production is then due only to the matter creation. It
is obvious that, when $\Psi = 0$, we shall expect that the
creation pressure vanish and so also the entropy production. We
shall express this by admitting a kind of phenomenological ansatz

\begin{equation}
\Pi =  - \alpha \frac{\Psi}{\Theta} ,
\end{equation}
where $\alpha$ is positive, so as to guarantee that, as argued
above, in the case $\Psi > 0$ and $\Theta > 0$ we shall have $\Pi
< 0$. With the choice (12), Eq. (9) is rewritten as
\begin{equation}
S^{\alpha}_{;\alpha} = \frac{\Psi}{T} (\alpha - \mu) ,
\end{equation}
or still, using Euler's relation,
\begin{equation}
S^{\alpha}_{;\alpha} = \Psi\sigma +(\alpha -  \frac{\rho + p}{n})
\frac{\Psi}{T}.
\end{equation}

It is convenient to compare (14) with the expression obtained by
covariant differentiation of (6), that is,
\begin{equation}
S^{\alpha}_{;\alpha} = \Psi \sigma + n\dot{\sigma } .
\end{equation}

It is immediate that
\begin{equation}
\dot{\sigma } = \frac{\Psi}{nT} (\alpha - \frac{\rho + p}{n}).
\end{equation}

\section{The Results of Prigogine et Alii}

The main results presented in Ref. [1] may be recovered if we
constrain the formulation of the previous section to the case in
which the specific entropy (per particle) $\sigma$ is constant
\begin{equation}
\dot{\sigma } = 0.
\end{equation}

Indeed, in this case, (16) implies that $\alpha = (\rho + p)/n$
and the creation pressure assumes the form

\begin{equation}
\Pi = - \frac{\rho + p}{n\Theta}\Psi ,
\end{equation}
which is exactly Eq. (13) of Ref. 1.

Notice now that Eq. (8) with condition (17) reduces to
\begin{equation}
\dot{\rho } = - 3H(\rho + p),
\end{equation}
of the FRW models. It must be clear, however, that (19) only holds
in the case $\sigma = const.$ Only then do $\rho$ and $n$
determine the equilibrium pressure $p$. In fact, from (19), we
have, for example, that
\begin{equation}
\rho = mn \Rightarrow p = 0;
\end{equation}
\begin{equation}
\rho  = aT4, n = bT3 \Rightarrow p = \rho / 3.
\end{equation}

In the general case, however, we obtain from (8), that
\begin{equation}
\dot{\rho } = \frac{\dot{n}}{n}(\rho + p) + nT\dot{\sigma},
\end{equation}
and $\dot{\sigma}$, or equivalently (in our formulation) $\alpha$,
must also be known for the pressure to be fixed.

From condition (17), we still obtain that

\begin{equation}
\frac{\dot{S}}{S} = \frac{\dot{N}}{N} = \frac{\Psi}{n}  ,
\end{equation}
where $S$ is the total entropy and $N$ the total number of
particles. Since $S$ and $n$ are positive, and the total entropy
cannot decrease, we must have $\Psi \geq 0$ and, in this case, as
the authors of Ref. 1 have concluded, the space-time can only
create matter, the reverse process being thermodynamically
forbidden. Nevertheless, this result only holds if $\dot{\sigma}
\leq 0$.  As a matter of fact, from (15), we see that the second
law of thermodynamics demands only
\begin{equation}
\Psi \geq - n\frac{\dot{\sigma}}{\sigma}  ,
\end{equation}
which is compatible with matter destruction ($\Psi < 0$) provided
$\dot{\sigma} > 0$.


\begin{thebibliography}{99}

\bibitem{1} Prigogine I., G\'eh\'eniau J., Gunzig E. and Nardone P., Gen.
Relativ. Gravit. 21, 767 (1989).

\bibitem{2} Bondi H. and Gold T., Mon. Not. R. Astron. Soc. 108, 252
(1948); Hoyle F.,  Mon. Not. R. Astron. Soc. 108, 372 (1948);
Hoyle F.,  Mon. Not. R. Astron. Soc. 109, 365 (1949);

\bibitem{3}  Dirac P. A. M., Nature 139, 323 (1937); Dicke R. H., Phys. Rev.
125, 2163, (1962); Canuto V., Adams P. J., Hsieh S. H. and Psiang
E., Phys. Rev. D 16, 1643 (1977).

\bibitem{4}  Wesson P. S., Astron. Astrophys. 119, 145 (1983); Wesson p. S.,
Gen. Relativ. Gravit. 16, 193 (1984).

\bibitem{5}  Tryon E. P., Nature 246, 396 (1973); see also Stenger V. J.,
Eur. J. Phys. 11, 236 (1990).

\bibitem{6}  Weinberg S., Astrophys. J. 168, 175 (1971); Udey N. and Israel
W., Mon. Not. R, Astron. Soc. 199, 1137 (1982); Lima J. A. S and
Waga I., Phys. Lett. A 144, 432 (1990).
\end{thebibliography}
\end{document}